\documentclass{PoS}
\usepackage{amssymb,amsmath,graphicx}

\title{Probing TeV scale physics in precision UCN decays}

\ShortTitle{Probing TeV scale physics in precision UCN decays}

\author{\speaker{Rajan Gupta} \footnote{LA-UR-13-29020} \\ 
	Theoretical Division, Los Alamos National Laboratory, Los Alamos, NM 87545, USA \\
	E-mail:\email{rajan@lanl.gov}}

\author{{Tanmoy Bhattacharya}\\
	Theoretical Division, Los Alamos National Laboratory, Los Alamos, NM 87545, USA \\
	E-mail:\email{tanmoy@lanl.gov}}

\author{{Anosh Joseph}\\
	Theoretical Division, Los Alamos National Laboratory, Los Alamos, NM 87545, USA \\
	E-mail:\email{anoshjoseph@gmail.com}}

\author{{Saul D. Cohen} \\
	Department of Physics, University of Washington, Seattle, WA 98195\\
	E-mail:\email{sdcohen@uw.edu}}

\author{{Huey-Wen Lin} \\
	Department of Physics, University of Washington, Seattle, WA 98195\\
	E-mail:\email{hwlin@phys.washington.edu}}

\abstract{We present the calculation of matrix elements of iso-vector
  scalar, axial and tensor charges between a neutron and a proton
  state on dynamical $N_f=2+1+1$ HISQ configurations generated by the
  MILC Collaboration using valence clover fermions.  These matrix elements
  are needed to probe novel scalar and tensor interactions in neutron
  beta-decay that can arise in extensions to the Standard Model at the
  TeV scale.  Results are presented at one value of the lattice
  spacing, $a=0.12$~fm, and two values of light quarks corresponding
  to $M_\pi=310$ and $220$ MeV. We discuss two sources of systematic
  errors, contribution of excited states to these matrix elements and
  the renormalization constants, and the efficacy of methods used to
  control them. }

\FullConference{31st International Symposium on Lattice Field Theory - LATTICE 2013\\
		July 29 - August 3, 2013\\
		Mainz, Germany}

\begin{document}

\section{Introduction}

The standard model (SM) of elementary particles has been very
successful in explaining phenomena up to the TeV scale and at the same
time the observed electroweak symmetry breaking in the SM points to
new physics at the TeV scale. In addition to the direct search for
novel particles and interactions, one can also look for their effects
in precision measurements at low energies.  In
Ref.~\cite{Bhattacharya:2011qm}, we showed that new scalar and tensor
interactions at the TeV scale could give rise to corrections at the
$10^{-3}$ level in precision measurements of the helicity flip parts
of the decay distribution of (ultra)cold neutrons (UCN). This
sensitivity is reachable in experiments currently under construction
and being planned. Even if these experiments see a signal, to
constrain the allowed parameter space of beyond the SM (BSM) models,
however, requires that matrix elements of isovector scalar and tensor
bilinear quark operators are known to 10--20\% accuracy. Lattice
calculations are well poised to provide these estimates with the
desired precision. In these proceedings, we summarize results on the
charges $g_A$, $g_S$ and $g_T$ calculated on 2+1+1 flavor HISQ
lattices~\cite{Bazavov:2012xda} using clover valence quarks at one
value of the lattice spacing, $a=0.12$~fm, and two values of light
quarks corresponding to $M_\pi=310$ and $200$ MeV. Details are given
in Ref.~\cite{Bhattacharya:2013ehc}.  We will also discuss the
efficacy of methods used to control two of the largest sources of
systematic errors -- contribution of excited states and estimates of
renormalization constants.

\section{Statistics}
\label{sec:stat}

The MILC Collaboration~\cite{Bazavov:2012xda} has generated ensembles
of roughly 5500 trajectories of 2+1+1-flavor HISQ lattices at three
values of light quark masses corresponding to $M_\pi=310$, 220, 140
MeV at $a=0.12$, $0.09$ and $0.06$ fm.  Here we focus on two ensembles
of roughly 1000 configurations at $a=0.12$fm with
$M_\pi=305.3(4)$ and $216.9(2)$ MeV~\cite{Bazavov:2012xda} (called
$M_\pi= 310$ and 220 MeV ensembles). These are separated by 5
trajectories of the hybrid Monte Carlo evolution and five hundred
trajectories are discarded for thermalization.  On each configuration,
we use four smeared sources, displaced both in time and space
directions to reduce correlations. To evaluate the statistical
significance of the data, we also analyze the data as two subsets with
roughly 500 configurations. These two subsets give compatible results
and the errors are roughly $\sqrt 2$ larger compared to the full
set. Our overall conclusions are: (i) the errors in $g_S$ are roughly
five times those in $g_A$ and $g_T$ and (ii) while statistics of
$O(1000)$ configurations provide estimates of $g_S$ with $15-20$\%
uncertainty, to achieve this desired accuracy after chiral and/or
continuum extrapoltions will require reducing the errors by at least
another factor of two, $i.e.$, increasing statistics by a factor of
$4-9$.

\section{Excited-State Contamination}

The goal is to extract all observables (charges, charge radii, form
factors) by calculating matrix elements between ground-state nucleons,
however, nucleon operators used on the lattice couple to the ground
state and all its radially excited states. The unwanted 
excited states contamination has to be removed to get the final estimate.

Assuming that only the leading excited state with mass $M_1$ and
coupling ${\cal A}_1$ to our operator contributes significantly, we
can write the three-point function with source at $t_i=0$, operator
insertion at $t=t$ and sink at $t_f= t_{\rm sep}$ as
\begin{align}
{\cal C}^{(3),T}_{\Gamma}(t_i,t,t_f;\vec{p}_i,\vec{p}_f) \approx \ &
       |{\cal A}_0|^2 \langle 0 | O_\Gamma | 0 \rangle  e^{-M_0 (t_f-t_i)} \ + \ 
       |{\cal A}_1|^2 \langle 1 | O_\Gamma | 1 \rangle  e^{-M_1 (t_f-t_i)} \nonumber\\
      +{} \ & {\cal A}_0{\cal A}_1^* \langle 0 | O_\Gamma | 1 \rangle  e^{-M_0 (t-t_i)} e^{-M_1 (t_f-t)} + {}\nonumber\\
      +{} \ & {\cal A}_0^*{\cal A}_1 \langle 1 | O_\Gamma | 0 \rangle  e^{-M_1 (t-t_i)} e^{-M_0 (t_f-t)} \,,
\label{eq:three-pt}
\end{align}
from which we need to extract $\langle 0 | O_\Gamma | 0 \rangle$. The
masses and amplitudes $M_0$, $M_1$, $A_0$, and $A_1$ are obtained from
the two-point functions. With these in hand, to extract $\langle 0 |
O_\Gamma | 0 \rangle$ by isolating $\langle 0 | O_\Gamma | 1\rangle $
and $\langle 1 | O_\Gamma | 1 \rangle$ requires that the calculations
be done with multiple $t$ and $t_{\rm sep}$. We have carried out
simulations at 5 values of $t_{\rm sep}$ for the $M_\pi=310$ MeV ensemble
and, based on insight gained from that analysis, on three values for
the $M_\pi=220$ MeV ensemble. Using the sequential source method, operator
insertion is carried out at all values of $t$ between the source and
sink timeslices.  We then apply a nonlinear least-square fitter that
automatically selects a fit range within $t_i-t_f$ for each $t_{\rm
  sep}$ value to reduce end effects and then fits data for all $t_{\rm
  sep}$ simultaneously using Eq.~(\ref{eq:three-pt}). 

\begin{figure}
\includegraphics[width=.99\textwidth]{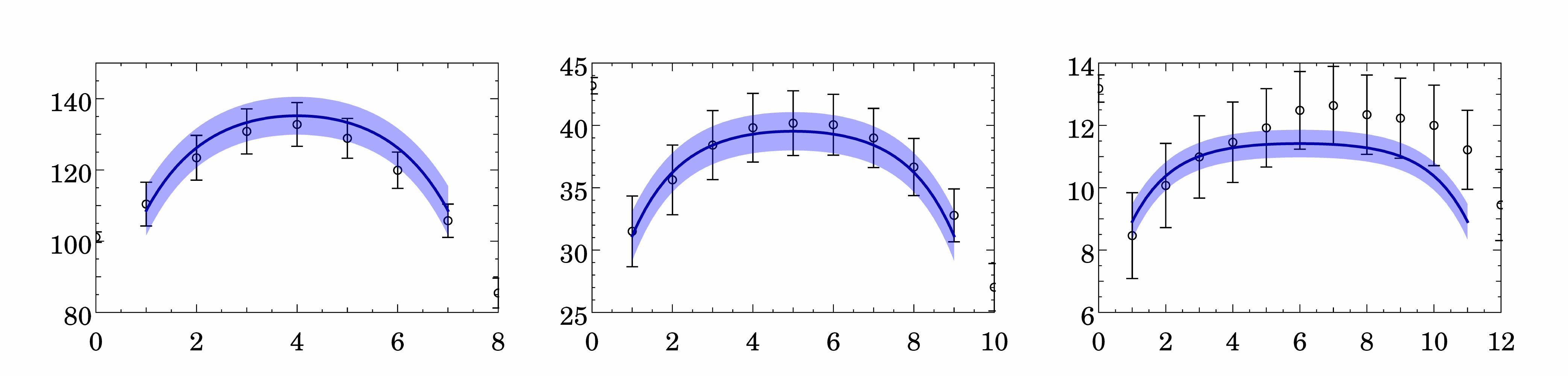} 
\vspace{-5pt}
\caption{Fit using Eq.(\protect\ref{eq:three-pt}) to extract $g_S$ from the
  220 MeV pion ensemble data at $a=0.12$fm. A simultaneous fit is made
  to the three $t_{\rm sep} = 8$, $10$ and $12$ data.  }
\label{Fig:22sim}
\end{figure}

Our analysis shows that excited state contamination,
contributions of non-zero $\langle 0 | O_\Gamma | 1\rangle $ and
$\langle 1 | O_\Gamma | 1 \rangle$, is significant and can be
eliminated by carrying out simulations at mutiple $t_{\rm sep}$ and
then doing a simultaneous fit to data at all $t_{\rm sep}$ using
Eq.~(\ref{eq:three-pt}).  An example of the simultaneous fit to $t_{\rm sep} = 8,\ 10,\ 12$ data 
to extract $g_S$ using Eq.(\ref{eq:three-pt}) for the 220 MeV pion ensemble 
at $a=0.12$fm is shown in Figure~\ref{Fig:22sim}. 
We also show a comparison of estimates obtained from 
different fit procedures in Figure~\ref{Fig:gFITS} and the efficacy of the 
2-state simultaneous fit method. Highlights of
our analysis for extracting the isovector charges $g_{A,S,T}$ are:

\begin{figure}
\includegraphics[width=.99\textwidth]{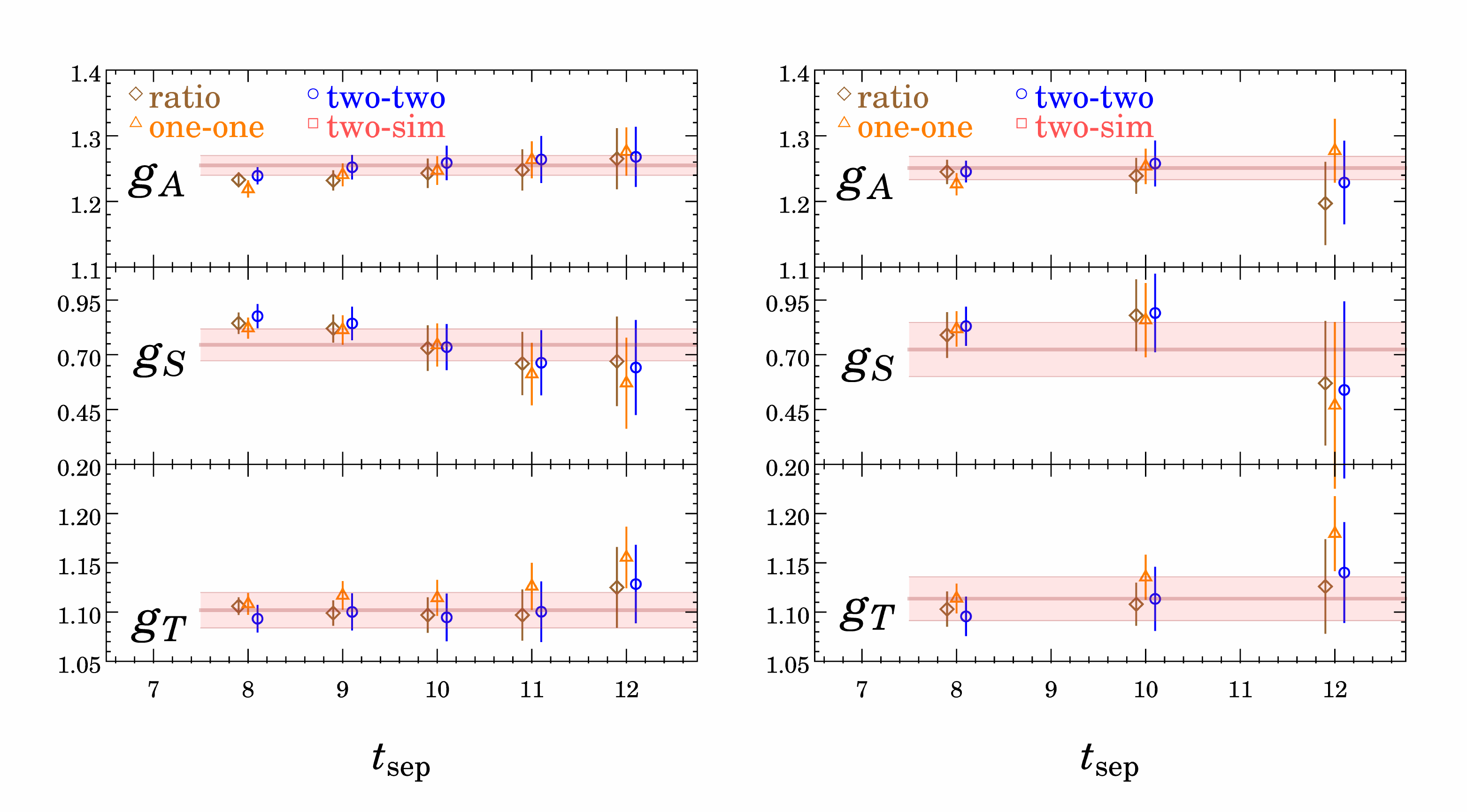}
\vspace{-5pt}
\caption{Comparison of estimates of the unrenormalized isovector
  charges $g_{A,S,T}$ as functions of $t_{\rm sep}$ with 310 MeV
  (left) and $220$ MeV (right) ensembles. The ``ratio'' and ``one-one'' method
  assumes a single state dominates; the ``two-two'' includes two states but
  analyzes data at each $t_{\rm sep}$ separately. The band gives the 
  result of the 2-state simultaneous fit to all $t_{\rm sep}$ (two-sim).  }
\label{Fig:gFITS}
\end{figure}

\begin{itemize}
\item The statistical errors increase by about $40$\% with each unit
  increase in $t_{\rm sep}$. This growth in errors limits the maximum
  $t_{\rm sep}$ that can be analyzed reliably with given statistics.

\item The 2-state simultaneous fit gives stable estimates of the
  central values and errors with respect to the range of $t$ selected
  for each $t_{\rm sep}$.

\item For the nucleon operator used by us, estimates from a single
  state ansatz becomes consistent wth those from the 2-state
  simultaneous fit for $t_{\rm sep} > 1.2$ fm.

\item The signal in the extraction of $g_S$ is the noisiest and the
  errors are about five times those in $g_A$ and $g_T$. Nevertheless,
  on the 220-MeV ensembles, the error estimate is about $15\%$,
  reasonably close to our desired accuracy.

\item The errors increase by about 20\% on lowering the light ($u$ and
  $d$) quark masses by a factor of two, $i.e.$, going from 310 to
  220 MeV ensemble. Unfortunately, the estimates at the two quark
  masses overlap within $1\sigma$ errors, therefore a reliable chiral
  extrapolation cannot be made. Based on current data we conclude that
  the best strategy is to work directly at the physical light quark
  masses, especially since the theoretical analysis of the expected
  chiral behavior of these charges is not well-established.
  
\end{itemize}

We are in the process of performing the same analysis on $a=0.12$,
$0.09$ and $0.06$ fm lattices at roughly the same light quark masses
corresponding to pion masses of 310 and 220 MeV. There is significant
improvement in the quality of the signal with decreasing lattice
spacing. With data at multiple values of quark masses and lattice
spacings in hand, we hope to elucidate the behavior versus quark
masses and make the extrapolation to the continuum limit.

\section{Non-perturbative Renormalization}

\begin{figure}
\includegraphics[width=.48\textwidth]{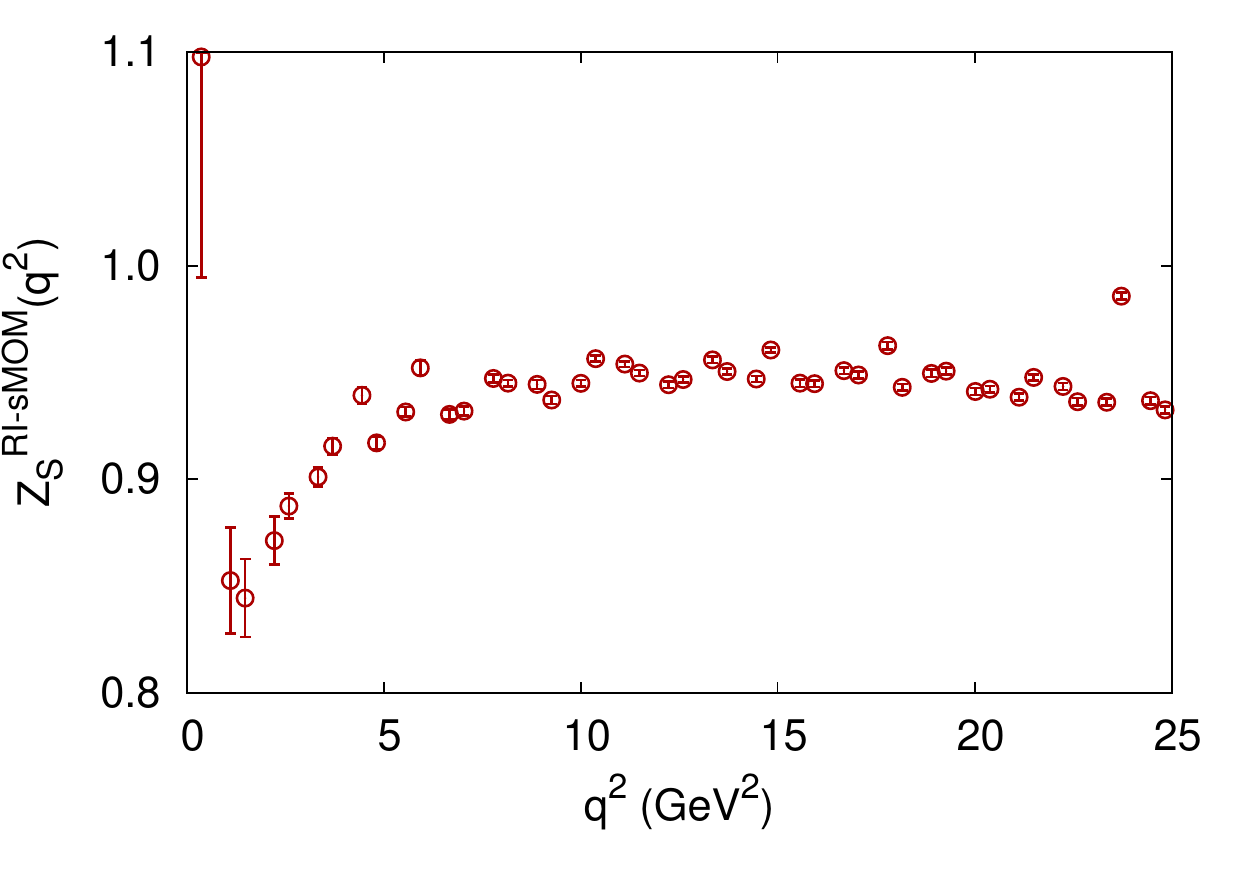}
\includegraphics[width=.48\textwidth]{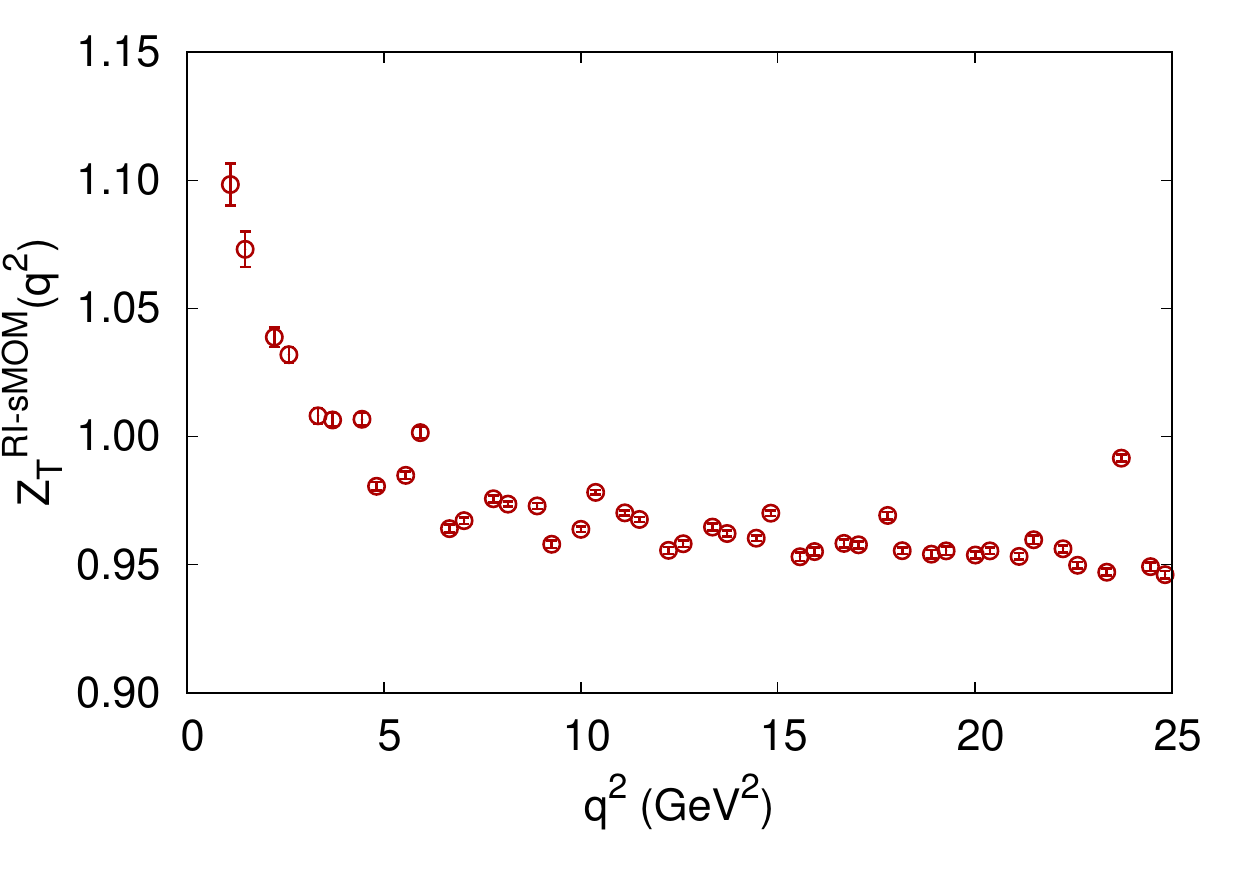} \\
\includegraphics[width=.48\textwidth]{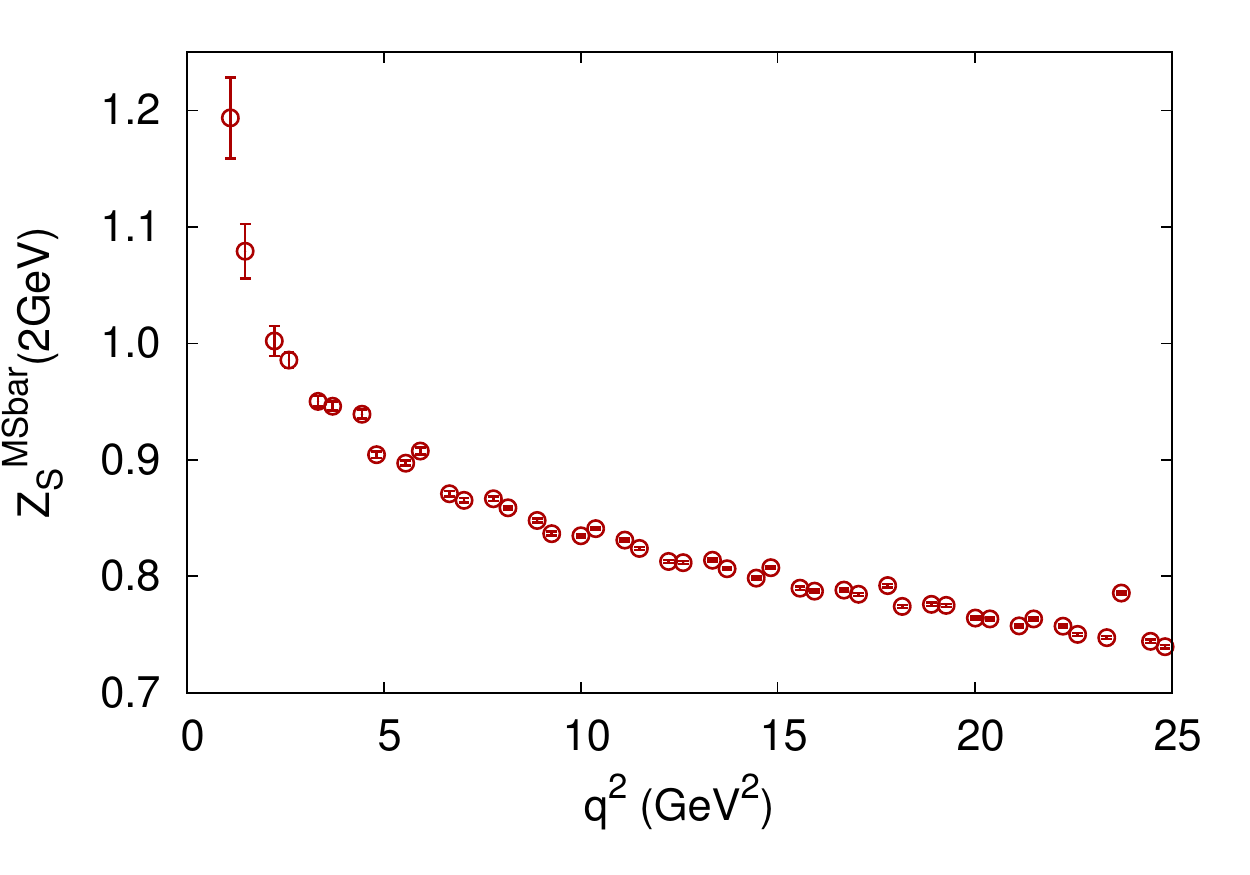}
\includegraphics[width=.48\textwidth]{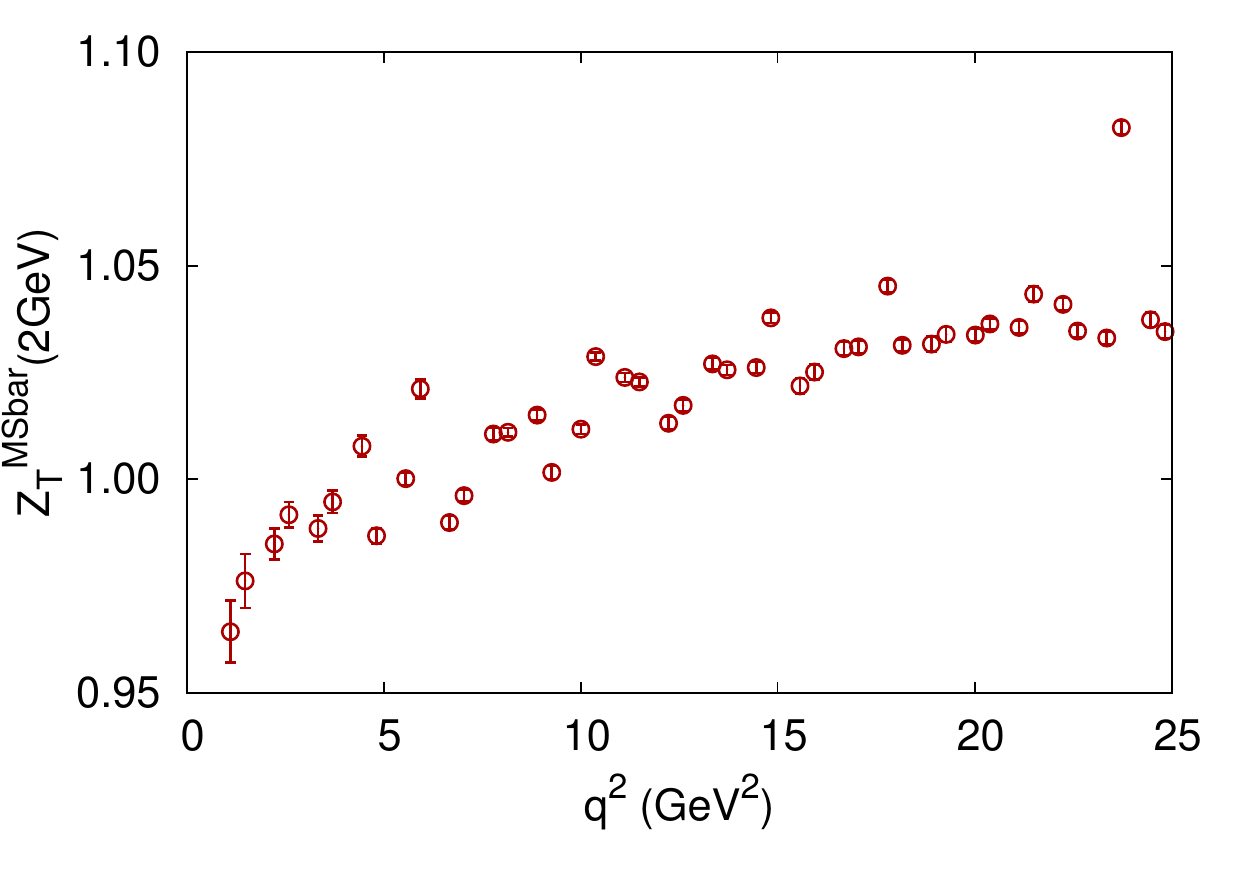} 
\vspace{-5pt}
\caption{$Z_S$ and $Z_T$ in the RI-sMOM
  scheme (top) and after running to 2 GeV in the $\overline{\rm MS}$
  scheme (bottom) for the $310$ MeV ensemble at
  $a=0.12$ fm.  The $q^2$ is the lattice momentum in physical units and
  at each $q^2$ we have plotted the $Z$ with smallest $\sum
  q_i^4/(q^2)^2$,$i.e.$, least breaking of rotational symmetry.}
\label{fig:Z-ST}
\end{figure}

We are using the RI-sMOM scheme to calculate the renormalization
constants of the isovector bilinears non-perturbatively on the
lattice~\cite{Martinelli:1994ty}. This method relies on the presence of
a window in momentum $q$, $ \Lambda_{QCD} < q < c/a$, where the lattice
artifacts are small and $c$ is a number of $O(1)$ that is {\it a priori}
unknown. Estimates in this window in the RI-sMOM scheme are matched to
the $\overline{\rm MS}$ scheme at the same scale $q$ using 1-loop
matching and then run to $2$ GeV using 2-loop expressions. Our results
in both the RI-sMOM scheme and $\overline{\rm MS}$ scheme at 2 GeV are
shown in Figure~\ref{fig:Z-ST} for $Z_S$ and $Z_T$ for the 310 MeV
pion ensemble at $a=0.12$ fm.

To interpret these results, it is important to establish the above
mentioned window in $q^2$. Sufficiently close to the continuum limit,
$Z_S$ and $Z_T$ in the RI-sMOM scheme should show a $q^2$ dependence
given by the anomalous dimensions of these operators and a weaker
dependence on the running of $\alpha_s$, $i.e.$, this known $q^2$
dependence can be used to establish a scaling window. For values of $q^2$
in such a scaling window, the results after conversion to
$\overline{\rm MS}$ scheme at $2$ GeV should be independent of
$q^2$. It is not obvious from the data shown in Figure~\ref{fig:Z-ST}
that such a window exists in our $a=0.12$ fm data. Data suggest that
the $a=0.12$ fm ensembles maybe too coarse to make contact with
perturbation theory. We see a marked improvement on the $a=0.09$ and $0.06$ fm 
lattices.

A second possibility is that the artifacts introduced by HYP smearing
of the lattice we use are large. Smearing is supposed to leave the
long distance non-perturbative physics unchanged but smooth out the
short distance fluctuations, $i.e.$, it introduces artifacts at large
$q^2$. Since smearing is a black box, it could shrink or completely
obscure the scaling window for a given quantity as discussed in
Refs.~\cite{smearing}.  In such cases one could take the following
approach. Assume that the calculation of the $Z's$ in RI-sMOM scheme has
lattice artifacts that persist when converted to $\overline{\rm MS}$
scheme at $2$ GeV. Then, for fixed $q^2$ and quark masses in physical
units we could calculate say $Z_S(\overline{\rm MS}, 2 {\rm GeV})\ g_S$ at
  different lattice spacings and extrapolate these estimates to the
  continuum limit. If all the significant systematics in the
  extrapolation to the continuum limit are well represented by
  extrapolation ansatz then in the continuum limit the results should
  be independent of $q^2$. These issues are currently being
  investigated using the data at all three lattice spacings.

\section{Impacts of $g_{S,T}$ on Searches for New Physics}

Precision measurements of neutron (nuclear) beta-decay and deviations
from well estimated standard model predictions would give hints of
potential BSM physics at the TeV scale. We can analyze new physics in
terms of an effective neutron beta-decay Hamiltonian:
\begin{equation}
H_{\rm eff} = G_F \left( J_{V-A}^{\rm lept} \times J_{V-A}^{\rm quark} + \sum_i \varepsilon_i^{\rm BSM} \hat{O}_i^{\rm lept} \times \hat{O}_i^{\rm quark} \right),
\end{equation}
where $G_F$ is the Fermi constant, $J_{V-A}$ are the left-handed
weak currents, and operators $\hat{O}_i$ have novel
chiral structure.  The low-energy couplings $\varepsilon$ encode both
the fundamental couplings at the TeV scale and their evolution to the
hadronic scale relevant to neutron decay.  As discussed in
Ref.~\cite{Bhattacharya:2011qm}, in neutron beta-decay only the 
isovector scalar and tensor operators contribute to $\hat{O}_i$. 

Novel scalar and tensor interactions can
also be probed at the LHC by analyzing the transverse mass
distribution in the channel $p + p \to e^{-} + \overline{\nu}_e +
X$. There, to distinguish novel from SM contributions, one
has to look at the distribution much above the $W$
resonance~\cite{Bhattacharya:2011qm}; the predicted bounds from
LHC become tighter with increasing center-of-mass energy and
integrated luminosity.

\begin{figure}
\begin{center}
\includegraphics[width=.48\textwidth]{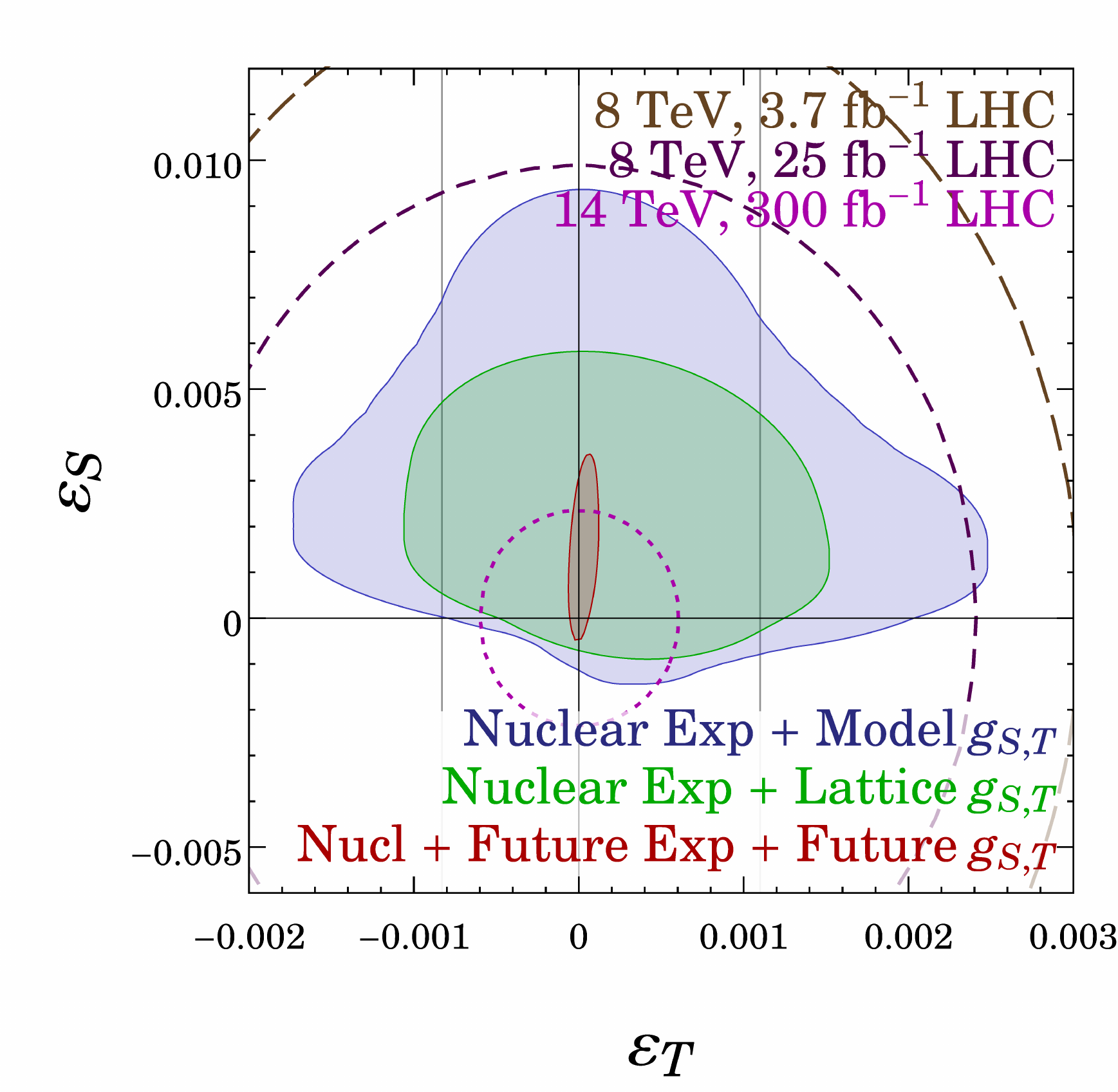}
\caption{
$\varepsilon_{S}$-$\varepsilon_{T}$ allowed parameter region using different
experimental and theoretical inputs as discussed in the text.  All
estimates are in the $\overline{\rm MS}$ scheme at 2~GeV.}
\label{fig:eSeT}
\end{center}
\end{figure}

Figure~\ref{fig:eSeT} shows an illustrative comparison of the constraints on
$\varepsilon_{S,T}$ (defined at 2~GeV in the $\overline{\rm MS}$
scheme) obtained from both low-energy
neutron decay and the CMS and ATLAS experiments at the LHC. We show
three bounds from the LHC for different center-of-mass energies and
integrated luminosity. To obtain these projected limits from the LHC,
we use the tail of the transverse-mass distribution in the reaction
$pp \to e\bar{\nu}+X$; that is, the region where $m_T > m_T^{\rm cut}$. The
transverse-mass cut is chosen such that the expected SM background
is less than one event. For the brown ellipse, the background is taken from
the measured value at CMS~\cite{CMS-PAS-EXO-12-010}; otherwise, the background
is estimated by computing at tree level the transverse-mass distribution due
to the production of a high-$p_T$ lepton from an off-shell $W$. For further
details of this analysis, refer to Refs.~\cite{Bhattacharya:2011qm,Bhattacharya:2013ehc}
and~\cite{Cirigliano:2012ab}. The outer dashed purple ellipse gives the LHC
expected constraint using the full current 8-TeV dataset; the inner dotted
magenta ellipse gives the expected final LHC constraint with maximum lifetime
luminosity at the 14-TeV design energy.

We compare these LHC constraints
to low-energy constraints using nuclear experiments.
The outer blue region
combines current nuclear experiments with model estimates of $g_{S,T}$
($0.25 < g_S < 1.0$ and $0.6 < g_T < 2.3$~\cite{Herczeg2001vk}).
The middle green region improves the constraint by using current lattice
values for $g_{S,T}$. The inner red region combines nuclear experiment with
anticipated future constraints from precision measurements of decays of
ultracold neutrons (assuming $|b_\nu-b| < 10^{-3}$ and $|b| < 10^{-3}$) and
future improvements in lattice values of $g_{S,T}$ to $10\%$ uncertainty.

We find that the eventual reach of low-energy and LHC constraints are
comparable.  For the LHC, this requires the full integrated luminosity of
300~fb${}^{-1}$ at 14-TeV center-of-mass energy, whereas for low-energy 
probes it requires that UCN experiments attain bounds better than $10^{-3}$ 
and $g_{S,T}$ are calculated with better than  $10\%$ error. 

\section*{Acknowledgments}
We thank the MILC Collaboration for sharing the 2+1+1 HISQ lattices
and Boram Yoon and Steve Sharpe for discussions.  Simulations 
were performed using the Chroma software suite~\cite{Edwards:2004sx} 
on facilities of the USQCD Collaboration 
funded by the U.S. DoE and Extreme Science and Engineering
Discovery Environment (XSEDE) supported by NSF grant number OCI-1053575.  RG is supported by DOE
grant DE-KA-1401020 and the LDRD program at LANL.

\end{document}